# Open Food Network: the Role of ICT to Support Regional Food Supply Chains in Australia

**Sherah Kurnia**
Computing and Information Systems
University of Melbourne, Australia
Email: sherahk@unimelb.edu.au

**Md Mahbubur Rahim**
Caulfield School of IT
Monash University, Australia
Email: md.mahbubur.rahim@monash.edu

**Patrice Braun**
Federation University
Australia
Email: p.braun@federation.edu.au

**Serenity Hill**
Open Food Foundation
Australia
Email: serenity@eaterprises.com.au

**Kirsten Larsen**
Open Food Foundation
Australia
Email: kirsten@eaterprises.com.au

**Danny Samson**
Management and Marketing
University of Melbourne, Australia
Email: dsamson@unimelb.edu.au

## Abstract

Many organizations have introduced various ICT-enabled innovations to improve economic, environmental and social performance. The Open Food Network (OFN) is an example of an ICT-enabled innovation that has the potential to enhance the sustainability of regional food supply chain by improving farmers' access to local and regional markets and consumers' access to fresh local produce, as well as optimizing the regional food distribution and improving local community welfare. OFN has just been recently launched in Australia and currently there is a limited understanding of the actual impacts. This research-in-progress paper aims to evaluate the effectiveness of the OFN system in connecting and supporting the sustainability of regional food supply chain communities in Australia that will help devise strategies for expanding the use beyond Australia. The findings contribute to a longer term research program that investigates how ICT can support sustainability initiatives within organizations and supply chains.

# 1  Introduction

In recent years, due to the rapid increase of global population and the significant economic growth in many developing countries, sustainability has become an important issue for businesses and the broader society (Porter and Kramer 2011). An increasing number of organisations have endeavoured to improve their business operations within their organisation and supply chains to enable them to achieve economic, environmental and social benefits. The various sustainability initiatives introduced have been based on the sustainable development principle (Brundtland 1987) that encourages the fulfilment of the needs of the current generation without compromising the ability of the future generations to meet their needs. Therefore, organizations are challenged to create not only economic benefits, but also environmental and social benefits. These three aspects (economy, environment and society) are the three pillars of sustainability which are known as the Triple Bottom Line (TBL) (Elkington 1997).

Specifically, within the food supply chain context, food waste has been identified as a significant supply chain issue due to growing supermarket domination in a number of countries (Blay-Palmer et al. 2013). Large supermarket chains around the globe typically apply high quality control which results in the rejection of imperfect produce supplied by local farmers. In addition, because of the restricted agreement on the supply imposed by large players within the industry, excess fruit and vegetables exists within supply chains and farmers have limited means of distributing these products (Blay-Palmer et al. 2013). Such a situation threatens the sustainability of regional supply chains that impact on the economic, environmental and social condition of the local communities in regional areas.

As a response to the current food supply chain issues, the Food Hub concept has emerged as a way to improve local food supply systems. Food Hubs represent an alternative model to the mainstream food supply and are defined as organisations that aggregate, distribute, and market food products primarily sourced from local and regional producers though simple, and more sustainable supply chains (Fisher et al. 2013). They help facilitate a closer connection between producers and consumers (Matson and Thayer 2013). The existing literature has identified a wide array of existing and potential benefits Food Hubs provide to producers, consumers, and local economies. Significant benefits include greater market access and the ability to provide a premium price through product differentiation by values associated with production, location, and farm identity. These benefits are particularly important for small and medium enterprises to seek alternative commodity markets to achieve economies of scale which are necessary for survival. Conner et al. (2008) also identify providing fresh food access to low-income communities, schoolchildren, and other institutions such as universities, prisons, and hospitals as an important benefit.

However, there are a number of key challenges experienced by Food Hubs that threaten their viability over a long term. For example, finding appropriate value chain partners for distributing products and developing mechanisms for value chain decision-making, transparency, and trust are difficult (Stevenson et al. 2011). In practice, many Food Hubs lack coordinated marketing mechanisms that raise brand awareness to local customers, and many have been unable to establish effective strategies for product differentiation, branding, and regional identity (Rose and Larsen 2013). Furthermore, Food Hubs have not been well recognized by government and mainstream players within the food industry (Rose and Larsen 2013).

With the advancement of the broadband internet technology, information and communication technology (ICT) has the potential to support the operations of Food Hubs and overcome some of the key challenges. Specifically, the application of electronic markets that was rapidly introduced in the late 1990s appears to be relevant to address the issues within local food supply chains. The value proposition of an electronic market is to reduce search costs for buyers and sellers through the use of a hub, facilitate product evaluation through transparency of information, and help discover the right price of products (Bakos 1998; Alt and Klein 2011). One of the recent innovative applications of e-market to support regional food supply chains is Open Food Network (OFN). Open Food Network has been established in the state of Victoria, Australia to connect various players of regional food supply chains in Australia and support the operations of Food Hubs. However, since such an application of an e-market is relatively new, there has been limited research conducted to assess the effectiveness of such an ICT-enabled innovation in supporting the sustainability and resilience of Food Hubs and the regional food systems. Furthermore, OFN has been recently launched in Australia and hence, little is known about the actual use and the impact of OFN on the stakeholders. Without an understanding of the effectiveness of OFN, it is difficult to devise appropriate strategies to enhance the features and encourage wider adoption within regional food supply chain parties.

This study is part of a large research project that aims to understand how ICT can help implement sustainability practices within organisations and supply chains. In this research-in-progress paper, we systematically assess how the OFN has been used by the early adopters who are regional food supply chain players and the impacts on their operations. The overall research question and sub-questions addressed in this study are:

*How does OFN support regional food supply chains in Australia?*
- How is OFN used by the adopting supply chain players?
- What are the benefits and challenges experienced by the OFN adopters?

This research-in-progress paper addresses the above research questions by conducting a preliminary focus group with nine current OFN enterprise users to investigate the use and impact of OFN. Through this preliminary investigation, we enhance the current understanding of how OFN as an example of simple e-Commerce application applied innovatively within the context of regional food supply chains can benefit the supply chain participants. In addition, we have identified a number of issues and challenges faced by the participants in using OFN that will be valuable for the OFN technology provider to further improve the OFN features and devise appropriate strategies that may encourage wider adoption. Understanding obtained from this study may also benefit future design of ICT-enabled initiatives that address sustainability in different contexts.

In the next section, we provide a brief literature review on the concept of Food Hubs, roles of ICT to support sustainability initiatives and e-Marketplaces as an example of ICT application. We then describe the concept of OFN, the research method, and the preliminary findings of the study. Finally, we draw some implications from this study and explain the next step of the research project.

## 2 Literature review

### 2.1 Food Hubs: Benefits and Challenges

The growing interest on a healthier diet to avoid controllable diseases such as diabetes and obesity, and concern over the environmental impact of supply chains has given rise to the popularity of supporting the concept of 'localisation". Localisation of food supply chains simply means that food should be consumed as close to the point of origin as possible (Seyfang 2006; Barham et al. 2012). In general, the concept of localisation stems from the consumers' interest in understanding where and how their produce is grown to ensure their food has high nutritional value without damaging the environment (Seyfang 2006). The concept had led to the emergence of 'Food Hubs' in regional areas of a number of regions including the United States, United Kingdom, and Australia.

Food Hubs could be viewed as an intermediary that uses an innovative business model to directly connect small and medium-sized producers and local consumers. Their potential to improve local food supply chain coordination, sustainability, and resilience has attracted the global attention of practitioners and researchers (Woods et al. 2013). Existing research indicates that Food Hubs may offer a number of environmental, social and economic benefits to the food industry and communities. For example, Food Hubs support environment friendly production such as organic items that do not use harmful ingredients such as pesticides and chemicals (Flaccavento 2009; Stevenson 2009). They provide nutritious produce in their local community by shortening the delivery timeline through restructuring of the supply chain (Rose and Larsen 2013). They also facilitate civic agriculture by helping community members understand the origin of their food and its supply system (Lyson 2005). In addition, they help improve the health of low-income communities, school children, and other institutions (e.g. universities, prisons and hospitals) by providing them with greater fresh food access (Erlbaum et al. 2011; Conner et al. 2008). Furthermore, studies indicate that Food Hubs ensure equitable income for farmers and food system workers, fair prices for consumers (Flaccavento 2009; Matson and Thayer 2013), and create more jobs in the local economy to support the development and operations of Food Hubs and regional food distribution systems (Rose and Larsen 2013). Finally, Food Hubs potentially minimize fuel consumption and carbon emissions by using optimisation technology to improve the routing schedule and truck optimisation (Rose and Larsen 2013).

Despite the growing popularity of Food Hubs, their operations regularly encounter a range of challenges. Pricing is a key challenge that requires fair price assurance for producers that are also affordable for consumers (Fischer et al. 2013). Food Hubs struggle to manage growth since they are newly established entrepreneurs with limited experience and business skills (Fischer et al. 2013). As the volume of transactions increase, many have found it challenging to manage the rise in suppliers, buyers, and operational costs associated with the growth and many have identified balancing supply and demand as a challenge (Fischer et al. 2013). Meeting capital requirements including the cost of new infrastructure, other start-up costs, and distribution has also been identified as a challenge (Clancy and Ruhf 2010; Melone et al. 2010; Fischer et al. 2013). Finding appropriate value chain partners and developing mechanisms for value chain decision-making, transparency and trust, determining effective strategies for product differentiation, branding and regional identity, and determining appropriate strategies for product pricing based on understanding true cost structures have also been identified as challenges (Martinez et al. 2010; Tropp and Barham 2008).

Within the Australian context, Rose and Larsen (2013) identify specific barriers and obstacles faced by the development of local food economies as a whole in the southern Melbourne region of Australia. They include difficulty in securing capital for business expansion and growth, lack of recognition of the new sector from government/mainstream players, inability to represent themselves and advocate their needs for large food manufacturers who can lobby to influence regulations, lack of integrated food and agricultural policy, regulations around food safety that hamper innovation, lack of coordination amongst policy makers, lack of highly-skilled labour, and lack of coordinated marketing mechanisms that raise brand awareness.

## 2.2 Roles of ICT to support sustainability initiatives

Currently, the literature exploring the link between ICT roles and Food Hubs is scant, although some IT-related challenges in the sector have been identified. Previous studies have indicated the roles that ICT can play in supporting organisations to practice sustainability initiatives within the supply chains (Dao et al. 2011; Porter and Kramer 2011). These roles include automating, informating, and transforming and providing infrastructure to support organisational activities in such a way that sustainability goals can be met (Dao et al. 2011; Kurnia et al. 2012). Kurnia et al. (2012), in particular, synthesize a number of key practices along the three dimensions of sustainability and conceptualize how each practice is supported by different roles played by ICT based on Dao et al. (2011)'s classification. They find informating and infrastructure provision to be the two major roles in supporting various sustainability practices. They also highlight the significance of transformation role to further support value creation in the sustainability context although it has not been well explored in the current literature. However, there are currently limited studies with empirical evidence to explain how different roles of ICT support organisations to improve the three aspects of sustainability performance (Elliot 2011; Kurnia et al. 2012).

Furthermore, ICT has been useful to support information sharing and facilitate collaboration among different organisations, opening the potential to address one of the challenges related to collaborating and goal alignment among different parties involved in the operation of Food Hubs (Dewet and Jones 2001). However, it appears that Food Hubs often have limited ICT skills and therefore do not have a clear understanding of ICT requirements to support business operations (Jablonski et al. 2011). They then experience lack of technical assistance related to web and data management, organizational management, product development, and food safety knowledge and compliance (Day-Farnsworth et al. 2009). A survey study involving Food Hubs in 2013 indicates that technology is one of their top three challenges (Fischer et al. 2013). There is some evidence that information technology developments in food supply chains connected to traceability, efficiency in distribution, quality systems, market information, and product development, are also being adapted to shorter, localized food chains (Barham et al. 2012 and Matteson and Hunt 2012) although there is little research on specific interventions and impact. Therefore, more effort is required to better understand how ICT can be used to support Food Hubs to enable them to run their operations more effectively and efficiently and how this is specifically linked to environmental, social, and economic benefits.

## 2.3 Electronic Marketplace (EM)

A number of capacity limitations and challenges faced by Food Hubs identified in the existing studies may be overcome by introducing an online intermediary (EM) to connect the producers and consumers and assist with the aggregation and distribution of fresh produce. Although EM comes in varying forms, it can be broadly defined as a virtual place where buyers and sellers meet to exchange goods/services (Segev et al. 1999). There are many benefits incorporating EMs in a business model. From a buyer's perspective EMs have the capacity to reduce search costs by making product and pricing information accessible, which also raises competition among suppliers, resulting in lower prices (Bakos 1998; Soh, Markus and Goh 2006). Electronic Marketplaces also allow sellers to compete for a wider range of customers, who may previously have been unreachable, with lower customer acquisition and transaction costs (Mahadevan 2003). Additionally, EMs can also enable sophisticated price discrimination on the sellers' behalf (Bakos 1998).

Electronic Marketplaces have evolved within multiple industries with differing modes of operation (Mahadevan 2003). They can facilitate trade between businesses (B2B) or between businesses and consumers (B2C, C2B and C2C) (Grieger 2003). Mahadevan (2003) gives an illustration of this breadth through a selection of EM structures which have arisen in the B2B space, including: catalogue aggregators, consortia marketplaces, forward and reverse auction sites, and Trading Partner Networks (TPN) and exchanges. According to Grieger (2003), a number of traits can be used to characterise the spectrum of EMs, among which are the degree of buyer/seller orientation and whether the pricing mechanism is fixed or flexible.

From the perspective of buyer/seller orientations, OFN falls into the category of a 'neutral' EM, that involves a third party equally favouring buyers and sellers (Grieger 2003). These EMs are referred to as 'market makers' and their typical business model is to facilitate interactions between buyers and sellers, charging a fee for this service (Benjamin & Wigand, 1995). Being IT based and unrestrained by brick and mortar resources, market makers have strong potential to upscale (Kaplan and Sawhney 2000). However, one challenge facing market makers is that they must recruit a large volume of buyers before they can attract sellers, or vice versa, placing them in a 'chicken-and-egg' predicament (Grieger 2003).

Additionally, the interests of buyers are in direct conflict with those of sellers in regards to the level of pricing transparency (Soh et al. 2006). While buyers are attracted to high price transparency, sellers may be deterred from participating in an EM with high transparency. Flexible pricing mechanisms such as forward and backward auctions, and pricing transparency features such as direct comparisons alter the price discovery process and raise competition (Bakos 1999). Neutral EMs must make a strategic decision that satisfies both parties. Soh, Markus and Goh (2006) assert that EMs must compensate buyers/sellers with other sources of value if the pricing transparency policy goes against their interests. In a discussion of the failings of past EMs, previous research indicates inadequate revenue as a common downfall of electronic markets (Kaplan and Sawhney 2000; Grieger 2003). Establishing a pricing model which appeals to customers and also generates adequate revenue to sustain operations has proven difficult.

## 3   Open Food Network (OFN)

The core vision of the Open Food Network is for a diverse, transparent, and decentralised food system. It provides food enterprises such as Food Hubs and other organizations involved in regional food distributions with an online marketing platform and tools for aiding their operational and administration activities. This is designed to enable Food Hubs to take advantage of e-commerce opportunities, without needing a large amount of capital to invest in the software. To the customer, the OFN website is a collection of online stores, organised as a directory. Customers can browse profiles and stores to learn about where their food is coming from and place and pay for orders. The profiles and online stores are controlled by the enterprises.

Open Food Network is being used by a variety of enterprises as their shopfront and as a tool for managing back-end activities. Current adopters of OFN include farmers selling directly to customers, not-for-profit food coops, and Food Hubs with multiple suppliers, buying group customers, and commercial customers.  These early adopters provide a test to whether OFN can help facilitate financially viable, alternative business models. The functionalities that serve the different enterprises include reporting tools, inventory management and the capacity to integrate accounts with trading partners (such as suppliers and buying group customers). This makes cooperative relationships and trading partnerships easier to manage, and lowers the administration burden for hubs to collaborate, therefore reducing overall costs in the supply chain. Ultimately, OFN's goal is to create a website with functionality that allows enterprises to establish an online trading presence in a relatively short amount of time and at little cost.

While e-commerce has reduced barriers to entry into the food and grocery sector, the cost associated with establishing an online shopfront and an Enterprise Resource Planning system for small business is still prohibitive to many micro and small enterprises. Further, duplicating the development of this software across multiple small businesses made little sense.  The OFN is evolving to fill this need.  More information on specific features is available at openfoodnetwork.org/features.

## 4　Research method

This study aims to understand the role of ICT in supporting regional food supply chains in Australia. In particular, we focus on assessing the effectiveness of the OFN as an example of an ICT application that has been established to support Food Hubs and other regional food supply chain parties, as well as address the challenges they face. Focus group study was used in this study as a way to obtain preliminary in-depth, qualitative information from the early adopters of the OFN. This method is beneficial for this type of exploratory study that investigates how ICT supports regional food supply chains.

OFN was launched in June 2015 and there are currently still a limited number of adopters. The focus group involved seven users of OFN as outlined in Table 1. All participants are enterprises that act as 'hubs". The focus group session was conducted for over two hours in June 2015. The session was recorded and subsequently transcribed for analysis. All the key ideas were coded under relevant thematic headings (Miles and Huberman 1994) to explore OFN use, benefits, and challenges as perceived by the OFN adopters.

| Organisation | Type | Year in Operation | Number of Participants | Participant Number |
|---|---|---|---|---|
| A | Food Hub | 5 years | 2 | Participant 1, Participant 2 |
| B | Food Hub | In planning | 1 | Participant 3 |
| C | Food Hub | 1 year | 1 | Participant 4 |
| D | Farm-based hub | 3 years | 1 | Participant 5 |
| E | School-based hub | Pilot | 2 | Participant 6, Participant 7 |
| F | Food Hub | 2 years | 1 | Participant 8 |
| G | Food Hub | 3 years | 1 | Participant 9 |

*Table 1. Overview of Focus Group Participants*

## 5　Study Findings

The following section describes the study findings which are grouped into 3 categories: use of OFN, benefits of OFN to enterprise users, and issues with OFN to enterprise users.

### 5.1　Use of Open Food Network

The Focus Group findings indicate that Open Food Network is being incorporated into a diverse range of food enterprise models. Firstly, OFN has been used to *facilitate a more efficient and customer friendly way of taking and processing orders*. This particular use is specifically relevant for Food Hubs with complex operations. Food Hub enterprises typically aggregate produce from six nearby producers each week to be distributed to multiple buying groups and institutional customers. They accept unique customer orders, in contrast to fixed mixed vegetable boxes, representing a relatively high degree of operational complexity. Product availability information is taken from farmers weekly. This information can now be displayed on the OFN accordingly. Each buying group has a unique shopfront created within the hub's order cycle, allowing for customised shopfronts according to the preferences of each buying group, such as exclusively organic.

With the use of OFN, individual customers place orders through their respective group shopfronts and upon closure of the order cycle, the central Food Hub views the order totals aggregated across all buying groups, and place orders with farmers accordingly. Produce is then brought to a central packing location either by delivery from the farmer, or collection by the hub. Orders are packed by paid staff and then picked up and delivered to the buying group pick up locations by a member of each buying group.

For a number of participants who have been performing direct marketing of their meat products, OFN has been used as *a direct marketing tool to reach regional and metropolitan customers efficiently*. This enables the participating enterprises to obtain advance customer orders. Taking orders in advance (through OFN) of slaughtering and packing gives an accurate indication of demand, reducing the risk of over or under supplying. Customers are also given flexibility to buy a 'mixed box', with the option of adding 'extras'. This can secure purchases of the less popular cuts of meat while also offering the option to order additional specific products.

OFN has also been used as part of a project of the Rural City Council *to increase accessibility of fresh fruits and vegetables to the community*, with the ultimate goal of reducing obesity. Open Food Network has been used to sell produce from a local wholesaler to buying groups established in local schools.

## 5.2 Benefits of OFN to Enterprise Users

A number of benefits of OFN have been achieved by the early adopters participating in our study, which are discussed below. Benefits of Electronic Marketplaces that have been identified in different contexts within the literature appear to be relevant for our study context.

### 5.2.1 Expanding marketing channels

The ability to market online is valuable to enterprises as it gives them an independent means of taking their product to market and allows customers to shop conveniently. Adopting OFN's e-commerce functionality gives enterprises a substantial advantage over more traditional marketing channels (e.g. farm gate stalls, markets, email ordering) because it helps enterprises connect with the rising number of Australian consumers who prefer the convenience of e-commerce over the traditional marketing channels. This sentiment is echoed in the following excerpts by Participant 9:

*"That rapidity with which you can scale up from having no online presence to being a fully-fledged online store …. …. People have all kinds of expectations of the ways that they want to engage with a food business these days. If you don't have an online presence, it's fairly difficult."*

### 5.2.2 Ease of use

The ease of product trial and use is very important to the emerging Food Hub movement. Minimising the learning costs and transactional costs is critical as there are already many mental hurdles preventing customers from engaging with the innovation. OFN has ensured that its shopfront and checkout processes are familiar to customers and easy to use which benefit the current and potential adopters. For example, Participant 5 noted: "*My customers say it is really easy to use, they like it, it's really visually appealing*".

### 5.2.3 Administrative efficiency

Administrative activities involved in operating a Food Hub include establishing produce availability of suppliers, accepting orders from customers, and managing accounts payable/receivable. Open Food Network's reporting functionality allows users to easily view and interpret data relating to their trading activities. A number of the participants indicate that the usability OFN's reporting functionalities is excellent. It gathers data about their products, customers, and orders, and lets them display and interpret this data in different ways. OFN has made the task of managing orders less demanding, "*I've got so many hours back every week in my life. I don't need to respond with individual emails to every order*" (Participant 5). For multi-party trading arrangements, this is particularly valuable, as the tasks of manually tracking orders and accounts can be a huge burden on these, often under resourced organisations.

### 5.2.4 Operational efficiency

The operations of local Food Hub businesses are inherently complex due to the seasonality and perishability of products, the inconsistency and unpredictability of supply, working with multiple suppliers and having multiple incoming and outgoing logistics, and tracking payments from customers to suppliers. Achieving efficiency in operations as well as automating and standardising processes is especially important for Food Hubs that are low on human resources or rely on contributions from under skilled and often transient volunteers.

The OFN software enables Food Hubs to manage the operations complexities more efficiently, making these factors more manageable and bringing feasibility to their business models. Participants 1 and 2 were acutely aware of the challenges they faced in juggling their operations. "*We had a desperate need for some sort of software to help us organise ourselves*" (Participant 1). They were also impressed by the "phenomenal flexibility" of the system. It allowed them to change the way they articulated their problem and their model and still find a way to adapt the system to find a solution. Another hub manager explained the simplicity of incorporating a new supplier into the system. "*It can happen very quickly… all of a sudden they're on the order cycle and we're beginning to move their stuff*" (Participant 4). The flexibility of OFN is highly valued by users, as it can accommodate shifts in their operations and the elements of uncertainty which are inherent in their businesses.

The OFN's order cycle model encourages enterprises to accept and fulfil orders in a periodic, routine manner. This allows them to aggregate demand on a weekly, fortnightly, or longer basis leading to more efficient operations and logistics. The order cycle function is being used by Participant 5, who has two complementary order cycles, which is used to coordinate with butchering and packing activities. As will be discussed in the limitations section, shopping in this cyclical manner involves a learning curve for customers.

### 5.2.5 Visibility and access to new customers

Functioning as a network of interconnected enterprises, rather than in isolation, OFN brings multiple related benefits to users: 1) Reach target customer segments (conscious consumers in their locale); 2) Connect with and form symbiotic trading relationships with other enterprises who share aligned interests; 3) Strengthen their image or brand through association with the OFN's values (stronger together than alone).

Multiple enterprises spoke of having attracted new customers from their involvement and exposure on OFN: "*then the first week, wham I got ten new customers*" said Participant 5, and Participant 4 said "*we've had at least one or two producers and some customers come along because of some kind of contact or awareness of OFN*".

Enterprises using OFN are able to share in the OFN's brand equity and strong brand values. "*I think a lot of people out there, a lot of consumers, want that information. And it's an added value thing… So that is the big selling point (of OFN), for both people setting it up for their own organisation, and people coming in and buying from it*" (Participant 3).

### 5.2.6 Raising the profile of Food Hubs and alternative food distribution models

Open Food Network is housing and giving momentum to Australia's local food movement, a movement which is still in its infancy and is in need of promotion and substantiation. Amongst the public, government, and mainstream players, there is a lack of recognition of Food Hubs and their potential role in regional food systems. By raising the profile and credibility of these novel models, OFN is improving the receptivity of government funders, suppliers, and consumers to the innovation, which makes for a more amenable environment for emerging Food Hubs. A local council employee who had piloted a school based Food Hub explained: "*OFN enabled us to have conversations that we haven't had before*" (Participant 5).

### 5.2.7 Reducing transaction costs

The administrative demands of coordinating multi-party trading can be prohibitive for many organisations. Open Food Network has functions which facilitate coordination and ease the exchange between parties, making the formation of complex, multi-party trading arrangements feasible. One existing hub operator explained how the OFN technology had led them to consider a greater diversity of food distribution models: "*It's (expanded) our notion of how local food comes in to a central point and how it's distributed and goes out.*"…"*Just using the word hub, we're able to see a much bigger role, for our group, than just food in and food out to families via veggie bag systems*" (Participant 1). Participants 1 and 2 have since explored supplying through wholesale channels, bulk buying groups, and the local community house, who serves low income families, amplifying their impact.

## 5.3 Issues with OFN to Enterprise Users

Since the OFN has just been recently launched, there are still a number of issues encountered by the early adopters, which are discussed below. Recognizing and identifying these issues are critical for further enhancement of the OFN features and devising appropriate strategies for encouraging wider adoption among regional food supply chain players.

### 5.3.1 Subscription and advanced customer accounts

Presently, OFN cannot attach detailed information to a customer's account, restricting the ways that enterprise can use OFN for their customer relationship management. The development of more detailed and integrated customer accounts would allow for a number of added features directly requested by participants, including: managing subscription/membership within the OFN system (Participant 9), creating "member only" order cycles (Participant 5) and creating automatic 'standing orders' placed by subscribers (Participant 5).

It is common for Food Hubs and Community Supported Agriculture enterprises (CSAs) to operate under a membership or subscription model. Currently, OFN does not offer functionality for managing subscription or creating 'member only' shops, which was raised as a major concern among participants. "*My big one is that we run on a subscription model. So we have customers subscribe for a 12 week season, and so not having a subscription management system as part of it, it's the biggest limitation for us*" (Participant 9). To compensate, this hub manager had to operate an additional external subscription database. In the case of member only shops, the manager must manually flag and contact non-member customers who have placed an order.

An extension of this shortcoming is the inability to create a recurring order for customers who have subscribed, or paid upfront for a repetitive order (such as a weekly vegetable box delivery). Thus, a request was placed by an enterprise (Participant 9) for "standing orders for particular products, and having repeating orders" to prevent enterprises from needing to manually place orders for their committed subscribers each week. These limitations show that there is space for OFN to create customer accounts which contain membership status and recurring order information, and can be used to further streamline manual administrative tasks.

### 5.3.2  Restricted checkout (requiring a minimum order or membership of enterprise)

Participants discussed a number of scenarios where they would like to restrict customers from placing an order. This includes: if they are not a member, if they have not purchased one essential item, if they have not reached a minimum spending amount, or if they are located beyond a set distance from the hub. At the time of writing, the OFN cannot restrict checkout based on any of the above. The type of restriction required varied depending on the enterprise. One producer wanted all customers to purchase the mixed box in order to checkout, with the option of adding additional extras: "*I would prefer it's a minimum order as a box, rather than a dollar amount*" (Participant 5). For another hub, a set spending amount was more important. This would make it possible for them to manage margins "*That's something that I foresee could be an issue, for people like us. Feasibility of fixed costs and everything, so you need to have a certain tipping point.*" (Participant 3)

### 5.3.3  Lack of integrations with other software

OFN does not integrate with any other software. One participant declared that they wanted to be able to integrate between their labelling scales, accounting package, and their OFN account. This would allow her to perform multiple functions in the same place, saving her time. In addition to the significant efficiency gains to be achieved from coordinating OFN with other software products, it would also reduce the barriers for enterprises who are migrating their existing operations to OFN and improve the ease of uptake.

### 5.3.4  Cross platform performance and responsiveness

A number of participants raised concerns about how well OFN performed on devises such as smartphones and tablets. "*I think the biggest (issue facing customers) is the mobile and tablet thing.*" Participant 9. One enterprise had customers who used tablets and mobiles exclusively, as they didn't own computers. Another hub was trying to secure hospitality customers, who traditionally ordered by fax. For them, this transition would be smoothest if they could order on their smart phones in the restaurant. "*Moving online was sort of blowing them away a little bit. And they were like, well how will our chefs do that? And I was like, well they've got smart phones*" (Participant 6). While OFN is designed responsively, the current performance (speed) impedes streamlined use from smartphones and tablets.

### 5.3.5  Set up time for existing enterprises

Depending on the size and complexity of the enterprise, there is a significant time commitment associated with establishing an enterprise on OFN. This time is in data entry (creating profiles and products) and also familiarising oneself with the different functions and how the site works: "*working out the nuts and bolts, and ironing out the creases to start with, that's a very, very long process*" (Participant 3). This was raised as a major challenge and a process that enterprise users wanted simplified. Specifically, Participant 5 said, "*If I could have just imported a spreadsheet, and all my products would be there, that would have been nice.*" While this user did persevere, she stated that "*that initial setup becomes quite a mental hurdle and finding the time to populate the site was difficult.*"

### 5.3.6  Lack of enterprise capacity and need for a forum for knowledge sharing

Throughout the focus group participants raised issues which confronted their business, and were not directly related to the OFN's core functions. Participants recognised that there was a huge range of skills and resources that enterprises need to be successful in the space; "*when you get into this, you realise* *there's quite a big suite of skills and knowledge you need to run a local food economy and these people don't have that... and I wonder whether there is that reason why a lot start up and crash.*" (Participant 6)

One participant highlighted that the "*Open Food Network already is somewhat directly educating people and providing that kind of function, and it could be expanded upon, and sharing learnings in different ways. I don't know, I see a lot of potential for that*" (Participant 4). An example of such a resource which proved useful for one participant was a collection of Food Hub case studies, housed at foodhubs.org.au; "*I had never worked in this area before and I read that website quite a lot before I started working there, and it really helped*" (Participant 6). A further suggestion was the provision of "*Something like a development kit, for people starting up in this space, in other places*" (Participant 9). Or simply, "*Even something associated to the OFN website, whereby you could go on and visit the help section for people using it*" (Participant 8). Participant 4 found pricing to be a challenge, "*I don't know whether there's something that OFN can do to help us through it.*"

They also recognised that "*there's a huge amount of knowledge sitting in the people who work around OFN, it would be a shame not to share that and to share the learning*"(Participant 4). There was consensus that some kind of online forum for sharing knowledge and having discussions with other OFN users would be very valuable: "*We could have people learning about Food Hubs and having that dialogue about, what do you put in your constitution, what sort of structure did you set up, and also some technical stuff about OFN, like how do I do this or that*" (Participant 3). One participant said, "*I'd love to hear about how other people operate, and get little ideas from them. Just the little things, like putting recipes out. What people are doing and how you might be able to improve yours, or change it a little bit*" (Participant 6).

### 5.3.7 Product property and quality control (e.g. free range)

At the time of the focus group there was no way for hubs or producers to attach a property field to products, to share any third party certification or labelling which applied to their products. This information was also not apparent to customers in the shop. "*The main thing is just putting a filter for certified organics*" (Participant 4). Such product information has become important for customers who are generally becoming more health conscious.

## 6  Discussion and Future Study

Based on the preliminary focus group conducted to explore how OFN as an example of an ICT application to support sustainability practices within regional food supply chains in Australia, we have identified some evidence of how different roles of ICT are played out by OFN to provide infrastructure to support sustainability initiatives and to automate, informate, and transform business processes within Australian regional supply chains. In line with a number of previous studies (Dao et al. 2011; Kurnia et al. 2012; Elliot 2011; Dewet and Jones 2001; Jablonski et al. 2011; Barham et al. 2012 and Matteson), our study shows that OFN enables regional food supply chain parties to achieve economic, environmental and social benefits by playing different roles to support various sustainability practices, but the most significant roles played are informating and infrastructure provision.

Table 2 demonstrates a number of key practices along the three pillars of sustainability that have been enabled by OFN through various roles played as suggested in Kurnia et al. (2012). The first two key practices related to profit margin and cash flow in Kurnia et al. (2012) have been combined into 'enhancing operational efficiency' to suit the study context. The shaded rows indicate the relevance of the key practices and the ICT roles proposed in Kurnia et al. (2012) for this study.

In terms of the economic dimension of sustainability, our study shows that OFN provides a useful infrastructure for regional supply chain parties to improve business administration and operational efficiency, increase the numbers of suppliers and customers by improving visibility and expanding marketing channels efficiently. These improvements are achieved through the automation and increase access to relevant information that is required to manage supply chain operations and control the quality of products. As a result, Food Hubs and other enterprises involved in the regional food supply chains can enhance sales and improve customer satisfaction.

In terms of the environmental dimension, the findings also indicate that OFN enables the elimination of traditional marketing activities which lead to reductions in the use of paper and printing requirements, transportations to visit specific farms and markets, and the use of space for facilitating interactions among the players. In addition, with the timely availability of supply and demand information, enterprises involved in the logistics can optimize their routing schedule to maximize the truckloads and minimize fuel consumptions.

| Dimension | Key Practice | ICT Roles Demonstrated through OFN | | | |
|---|---|---|---|---|---|
| | | Automate | Informate | Transform | Infrastructure |
| Economy | Enhancing operational efficiency | Y | Y | | Y |
| | Achieving consumer satisfaction | | Y | | Y |
| | Creating repeat customers | | Y | | Y |
| | Enhancing sales | Y | Y | | Y |
| | Quality initiatives | | Y | | Y |
| | Creating competitive advantages | | Y | Y | Y |
| Environmental | Eco-design of products | Y | Y | Y | Y |
| | Green purchasing | Y | Y | Y | Y |
| | Clean/Lean Production | | Y | Y | Y |
| | Green Distribution | | Y | Y | Y |
| | Reverse Logistics | | Y | Y | Y |
| Social | Community relations | | Y | Y | Y |
| | Employee well-being | | Y | Y | Y |
| | Human rights | | Y | | Y |
| | Work safety/healthier community | | Y | | Y |
| | Ethical considerations | | Y | | Y |
| | Purchasing from minority-owned suppliers | Y | Y | | Y |
| | Product safety/quality | | Y | Y | Y |
| | Education support | | Y | | Y |

*Table 1. Relevant Sustainability Practices and the Roles of OFN (based on Kurnia et al. 2012)*

Finally, in terms of social benefits, the findings point towards useful evidence of how OFN through its roles in providing infrastructure, automation, informating and transformation can help generate some social benefits. For example, by playing those four roles, OFN affects the way food enterprises interact with suppliers and customers which in turn contributes to the community relations through online forums to exchange ideas and mutual supports. Likewise, OFN helps improve food enterprises' employee well-being through enhanced skills and greater operational efficiency and education support to the community. Through product quality and information transparency, OFN even potentially creates a healthier and more resilient community.

Besides the benefits of the OFN, this study has identified a number of challenges that still need to be addressed before we can expect a wider adoption of the OFN. The challenges, however, can be addressed through continuous improvement of the OFN features and ongoing education support to the regional communities to leverage the potential of ICT and to harness on knowledge sharing and collaboration within the communities to establish sustainable local food economy. This study confirms that OFN is a useful ICT tool for supporting sustainability initiatives within regional food supply chains and extends the current understanding of how ICT can support sustainability practices with organizations and supply chains. In the next step of our research, we will conduct a longitudinal study to find out the longer term impacts of the OFN on those early adopters and the overall sustainability practices within the same regional community. We will also do a comparative study with the UK and US experience as OFN is being rolled out globally. All this will contribute to the theory development regarding the importance and roles of ICT in supporting sustainability initiatives.